\documentclass{PoS}

\usepackage{amsmath}
\usepackage{amssymb}
\usepackage{mathrsfs}
\usepackage{graphicx}
\usepackage{cite}
\usepackage{mathpazo}
\usepackage{float}

\title{Sextet Model with Wilson Fermions}
\ShortTitle{Sextet Model with Wilson Fermions}
\author{
	\speaker{Martin Hansen} \\
	CP3-Origins, University of Southern Denmark, Campusvej 55, DK-5230 Odense M, Denmark \\
	E-mail: \email{hansen@cp3-origins.net}
}

\author{
	Claudio Pica \\
	CP3-Origins, University of Southern Denmark, Campusvej 55, DK-5230 Odense M, Denmark \\
	E-mail: \email{pica@cp3-origins.net}
}

\abstract{
We present new results from our ongoing study of the SU(3) sextet model with two flavors in the two-index symmetric representation of the gauge group. In the simulations use unimproved Wilson fermions to investigate the infrared properties of the model. We have previously presented results for the spectrum of the model \cite{Drach:2015sua} in the weak coupling regime. Here, to better understand the overall behavior of the lattice model, we map its non-trivial phase structure in the space of bare parameters. At strong coupling, we observe a first order phase transition when decreasing the bare quark mass. This first order transition weakens when moving towards weaker couplings with an endpoint at a finite value of the bare coupling, after which it appears to be a continuous transition. We also investigate the behavior of the mass spectrum and scale-setting observable, as a function of the quark mass, and show that their qualitative behavior change significantly when moving from the strong coupling into the weak coupling phase.
\\[2mm]
\it Preprint: CP3-Origins-2016-044 DNRF90
}

\FullConference{
	The 34th International Symposium on Lattice Field Theory \\
	24-30 July 2016 \\
	University of Southampton, UK
}

\begin{document}
\small

\vspace{-2mm}\section{Introduction}\vspace{-2mm}
In this work we discuss the phase structure of the SU(3) sextet model, when simulated on a lattice with unimproved Wilson fermions and the plaquette gauge action, and we investigate the behavior of the model in the weak and strong coupling phase.  

In the weak coupling region we have carried out a number of large volume runs to study the behavior of the spectrum in the chiral limit, as reported in \cite{Drach:2015sua}. While these simulations have been improved, at present our data is insufficient to clearly distinguish between conformality and chiral symmetry breaking.

In this proceeding, we first present our results for the phases of the lattice model in section~\ref{sect:pd} and finally in section~\ref{sect:b} we report on the behavior of some relevant observables for the model in the weak and strong coupling phases.
In particular we show that the behavior of the meson spectrum depends strongly on the bare coupling i.e. which phase the system is in.

\begin{figure}[t]
\centering
 \includegraphics[scale=0.85,trim={0 2mm 0 5mm},clip]{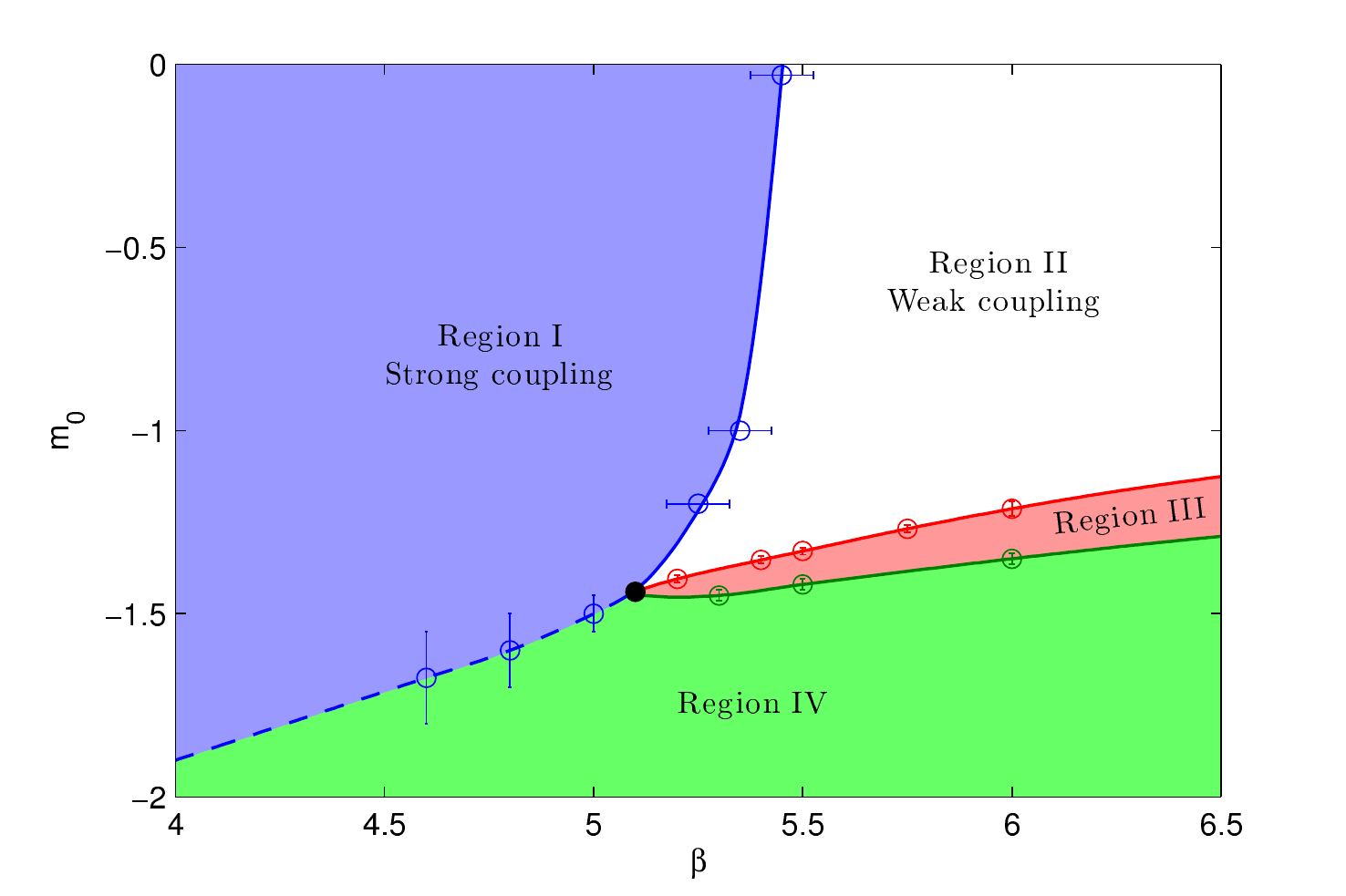}
 \caption{\footnotesize Phase diagram for the lattice model showing four different regions as a function of the bare parameters $\beta$ and $m_0$. Region I is a strong coupling bulk phase, Region II and III are weak coupling phases with positive and negative PCAC mass, respectively, and Region IV is an unphysical artefact region.}
 \label{fig:phase_diagram}
\end{figure}

\vspace{-2mm}\section{Phase structure\label{sect:pd}}\vspace{-2mm}
To properly understand the lattice model and reveal its non-trivial phase structure, we have performed an extensive scan (comprising more than 200 simulations) in the parameter space of the bare coupling $\beta$ and the bare quark mass $m_0$. For this scan we use either $8^4$ or $16^3\times32$ lattices, depending on the observable. In order to check for finite volume effects, some of the simulations have been repeated on $24^3\times48$ lattices.

In Fig.~\ref{fig:phase_diagram} we show the phase diagram as determined by our numerical simulations. We identify four different regions separated by first order (dashed in the figure) or continuous (solid) transition lines.   

A first continuous crossover transition line is identified (solid blue line) by looking at the peak of the plaquette susceptibility (as a function of $\beta$) in the region of positive PCAC mass. This line separates the ``weak coupling phase'' from the ``strong coupling phase''. The fact that the susceptibility is volume independent, indicates that this is a smooth crossover between the two
\begin{figure}[H]
\centering
 \includegraphics[scale=0.8,trim={4mm 20mm 0 18mm},clip]{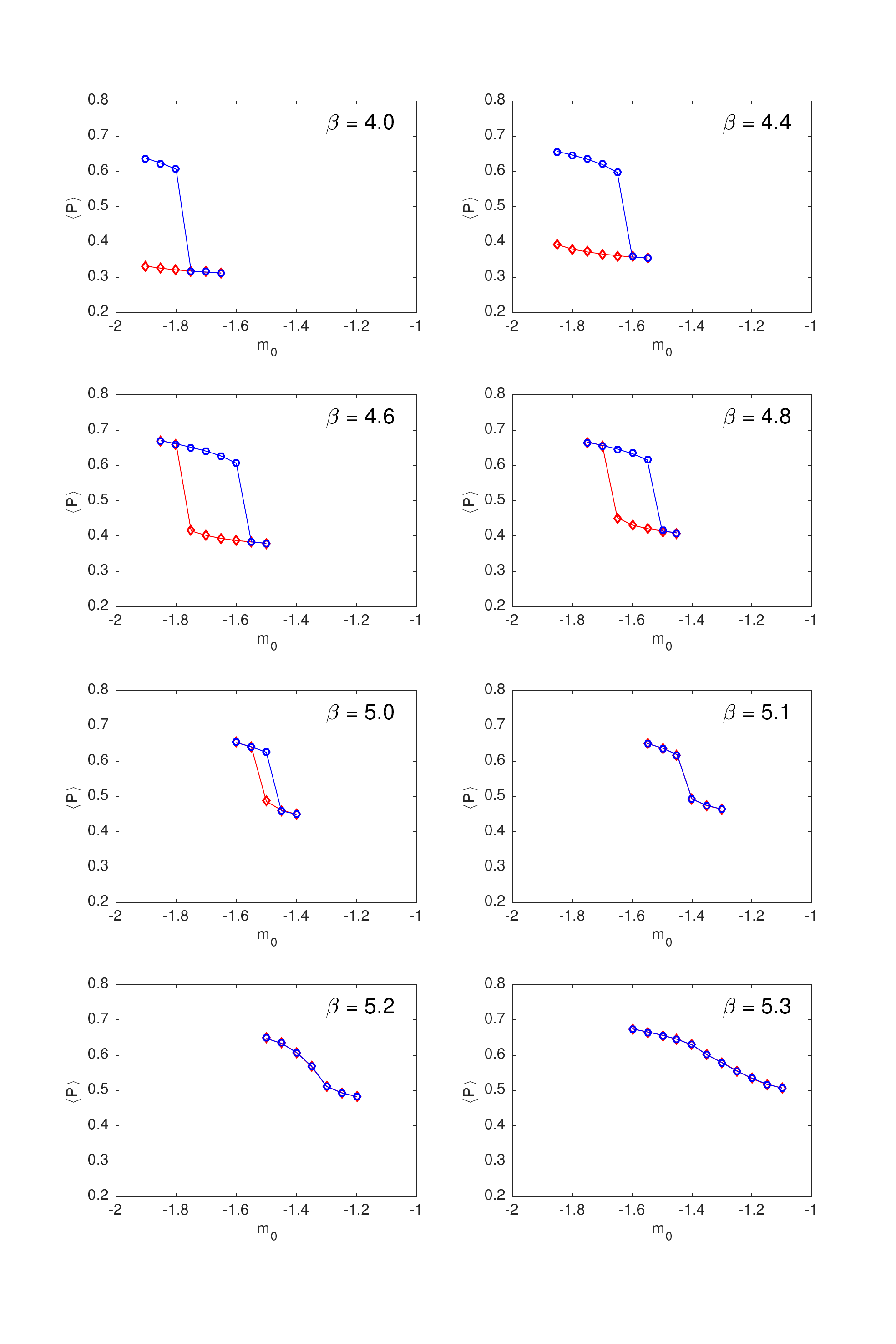}
 \caption{\footnotesize Hysteresis cycles around the first order phase transition. The red (blue) points indicate simulations started from a random (unit) configuration. At strong coupling we are unable to run simulations for bare masses below $m_0\approx-1.9$. As we approach the weak coupling phase, the first order transition disappears.}
 \label{fig:hysteresis}
\end{figure}
\noindent regions. We also note that this crossover seems to become sharper when moving to lower bare masses.

At strong coupling a line of first order transition is present (the dashed blue line) with an endpoint at $\beta\approx 5.2$. This transition line is identified from hysteresis cycles in the value of the plaquette, as shown on Fig.~\ref{fig:hysteresis}. Here all data points have been obtained from simulations started independently from either a unit configuration (blue points) or a random configuration (red points). At very strong coupling we are unable to run simulations for bare masses below $m_0\approx-1.9$, which is why the hysteresis cycles in top most panel are not closed. As we approach the weak coupling region, the hysteresis cycles disappear, as seen in the panel at the bottom. At $\beta=5.2$ one can still observe a small remnant of the transition, while at $\beta=5.3$ it is absent. For the study of hysteresis cycles, we used small $8^4$ lattices, since the presence of strong metastabilities on large lattices makes it difficult to perform numerical simulations across the transition. To check for the persistence of the first order transition, we have repeated the simulation at $\beta=5.0$ and $m_0=-1.5$ on the three volumes $8^4$, $16^3\times32$ and $24^3\times48$. In the hysteresis region, we have measured the PCAC mass which appears to have different signs in the two metastable states (see next section for more details).

The line of first order transition continues as a continuous transition line (solid red line) in the weak coupling regime. This line is identified as the line where the PCAC mass vanishes, when approaching from positive bare mass.

Finally one more continuous transition line is found (solid green line) at weak coupling, where the slope of the PCAC mass changes sign. Between these last two transition lines, the PCAC mass is negative (Region III in the plot).

The outlined phase diagram shares a number of features with the phase diagram for the SU(2) gauge theory with two Dirac fermions in the adjoint representation, including a first-order phase transition at strong coupling, together with an unphysical artefact region \cite{Hietanen:2008mr}. In simulations with many fundamental flavors, the first-order phase transition is also observed \cite{Nagai:2009ip}, indicating that this is a general feature for Wilson fermions when approaching the conformal window.

\vspace{-2mm}\section{Spectral and scale setting observables \label{sect:b}}\vspace{-2mm}
In this section we investigate the behavior of several observables of interest to understand the physical properties of the theory, namely the PCAC mass, the masses of the pseudoscalar and vector meson states and the scale setting observables $t_0$ and $w_0$.

\begin{figure}[t]
\centering
 \includegraphics[scale=0.56,trim={6mm 0 10mm 0},clip]{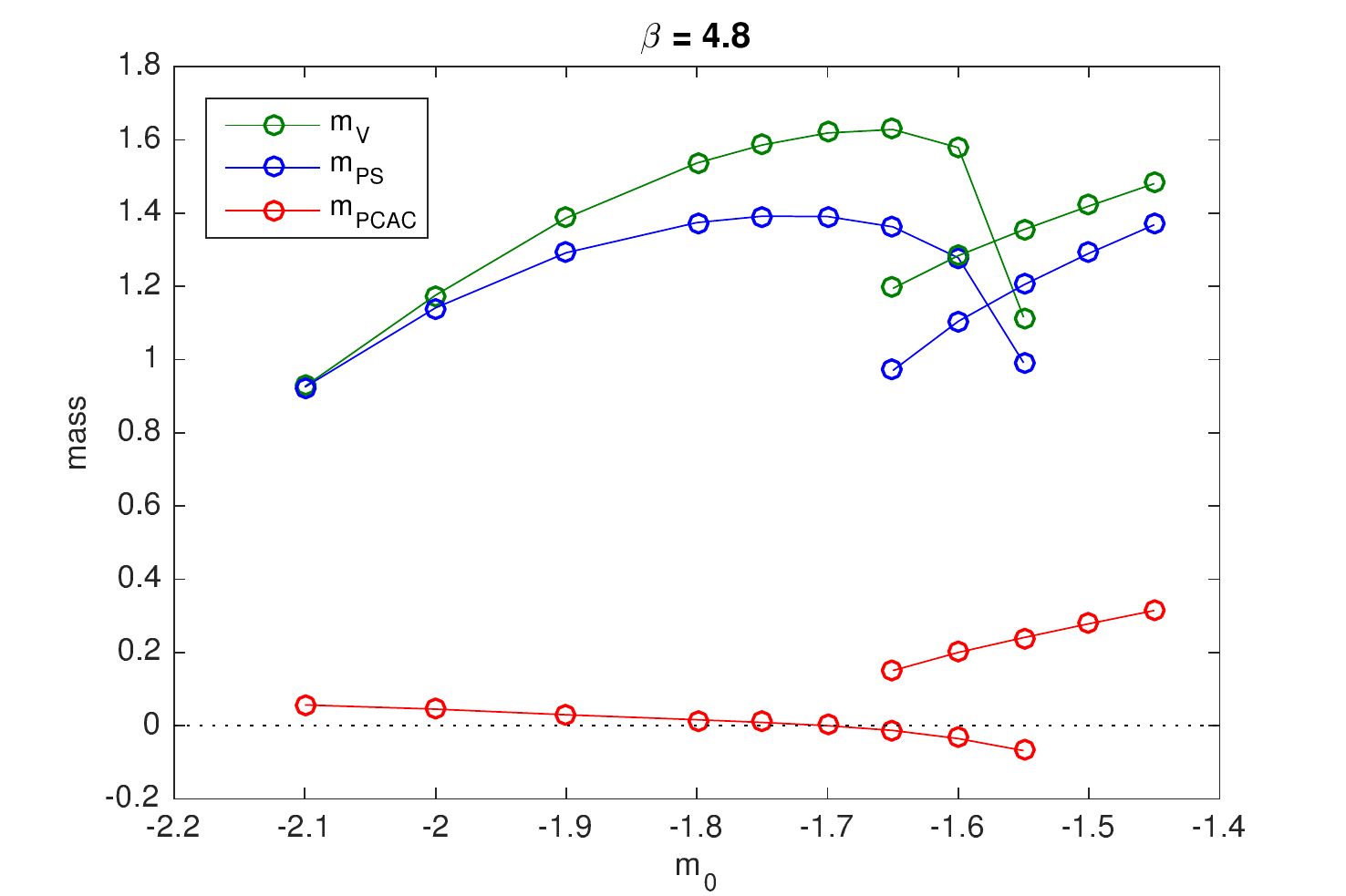}
 \includegraphics[scale=0.56,trim={6mm 0 10mm 0},clip]{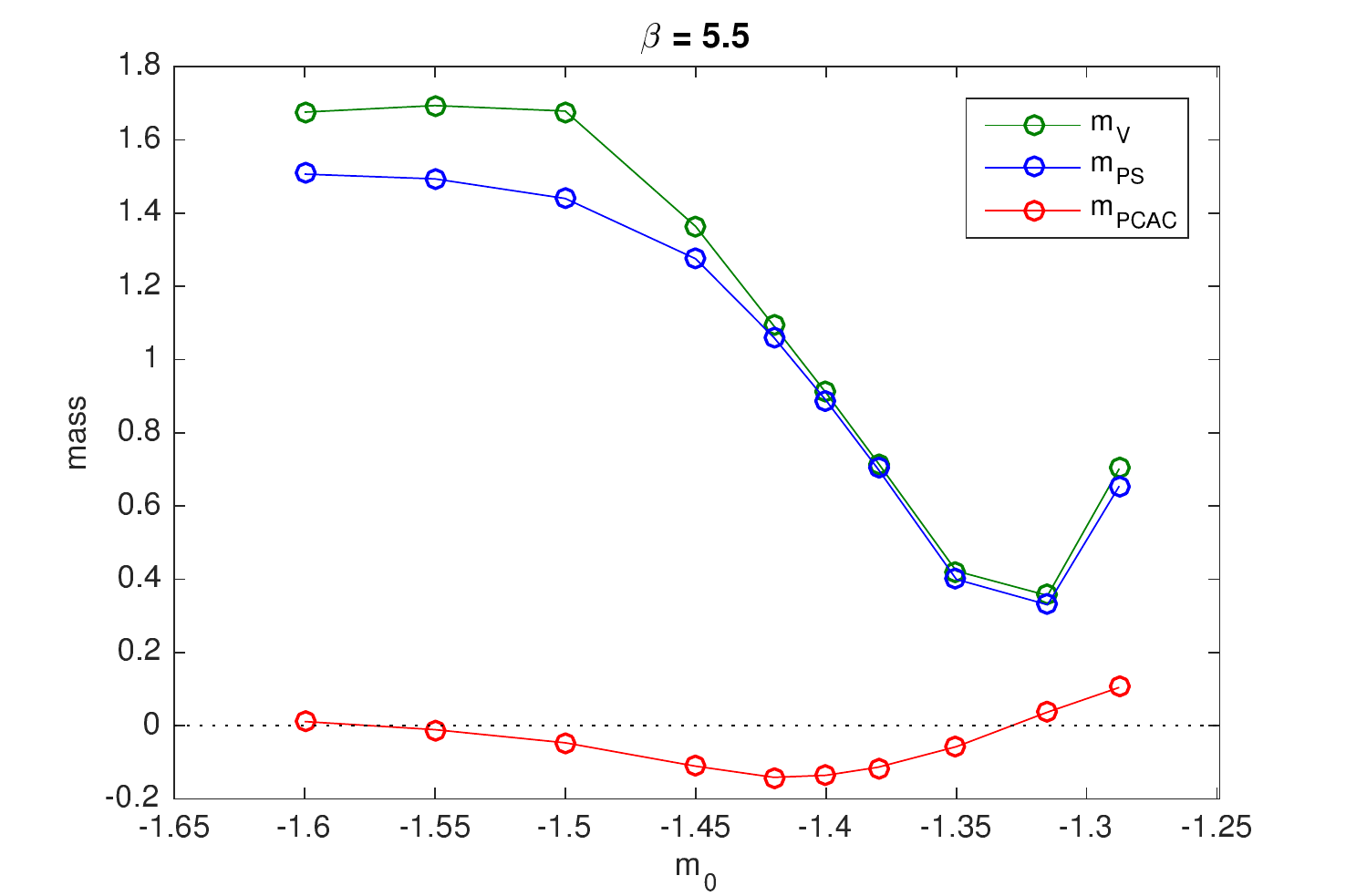}
 \caption{\footnotesize \textit{Left}: Behavior of $m_V$, $m_{PS}$, and $m_{PCAC}$ when crossing the first-order transition at $\beta=4.8$. The lines extending to the right (left) are simulations in Region I (Region IV). We observe that the slope of the PCAC mass has opposite sign in Region IV. \textit{Right}: At $\beta=5.5$ there is a continuous transition between Region II, III and IV, and we clearly see that the slope of the PCAC mass changes sign around $m_0\approx-1.42$.}
 \label{fig:1storder}
\end{figure}

\begin{figure}[t]
\centering
 \includegraphics[scale=0.56,trim={6mm 0 10mm 0},clip]{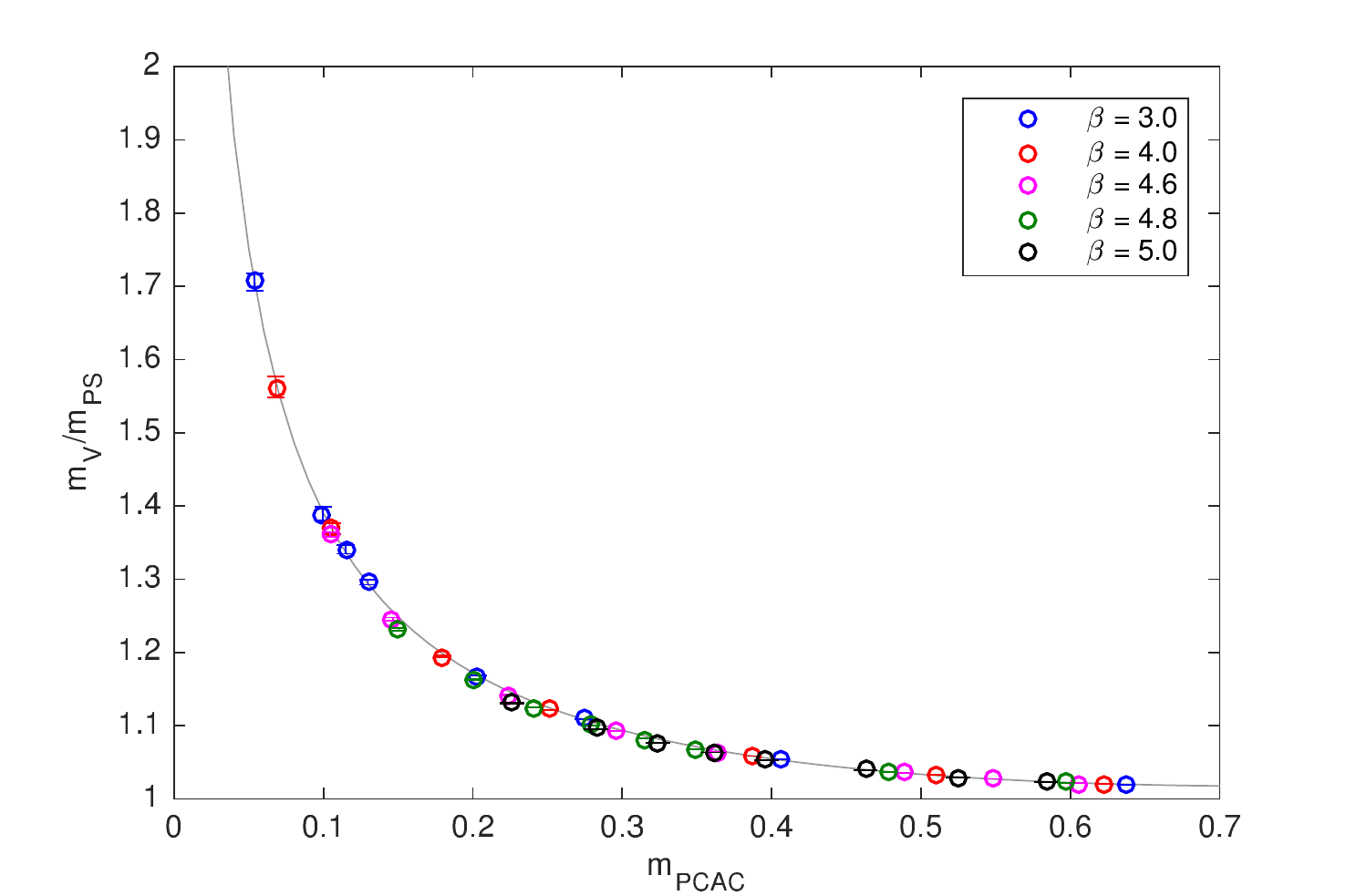}
 \includegraphics[scale=0.56,trim={6mm 0 10mm 0},clip]{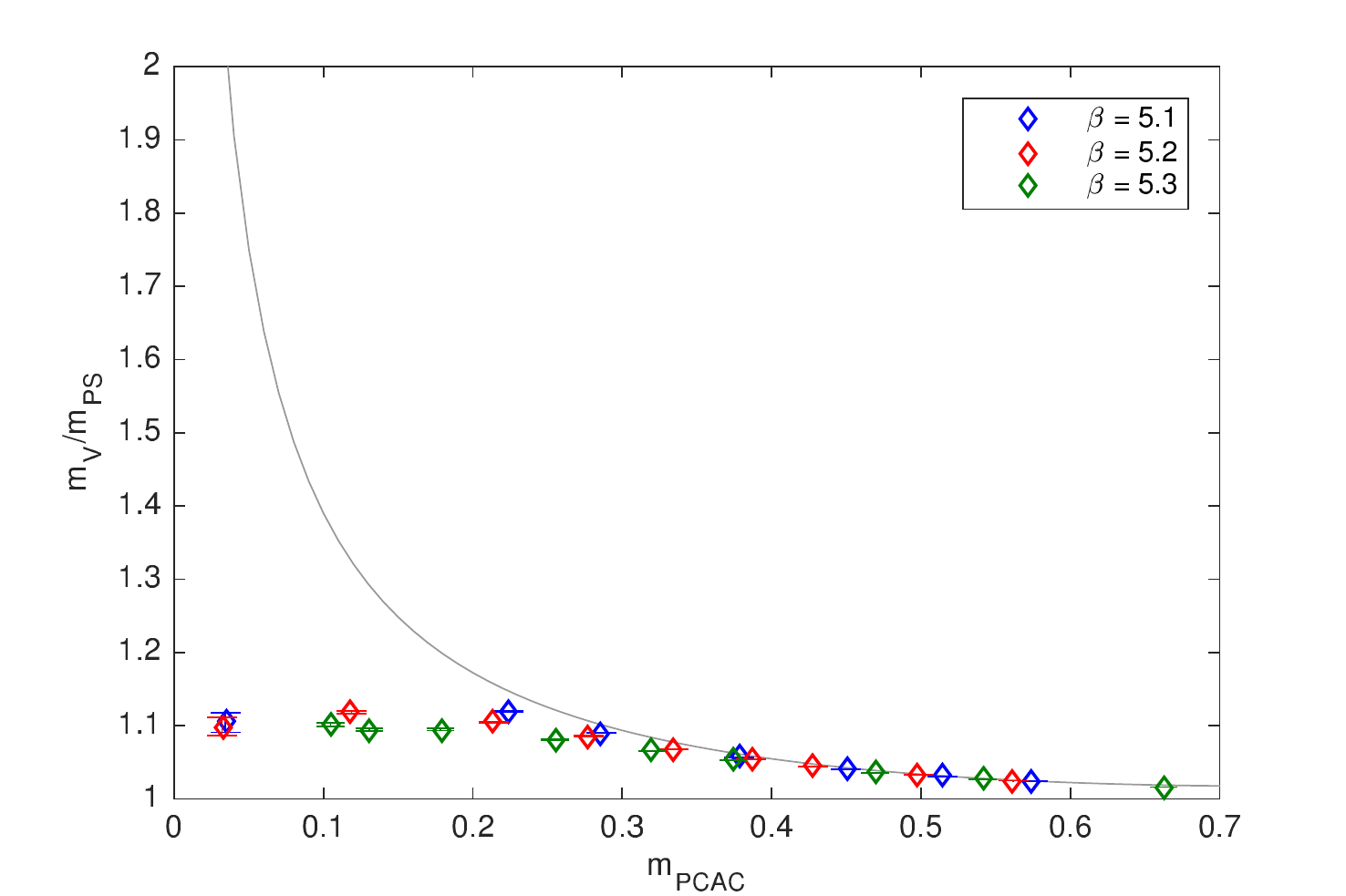}
 \caption{\footnotesize The ratio $m_V/m_{PS}$ as a function of the PCAC mass and inverse coupling $\beta$. \textit{Left}: Inside the bulk phase (Region I) the ratio diverges in the chiral limit, as expected from chiral symmetry breaking. The gray line is a fit to the data at $\beta=3.0$ and $\beta=4.0$. \textit{Right}: In the weak coupling phase (Region II) the ratio is constant in the chiral limit, as expected in a conformal model.}
 \label{fig:mvkmps}
\end{figure}

In Fig.~\ref{fig:1storder} we show how the PCAC mass, the pseudoscalar and vector masses change as a function of the bare quark mass for different lattice phases. At $\beta=4.8$ (left panel) a hysteresis region is visible, signalling the presence of the first-order transition, and we observe discontinuous jumps for the masses. In the figure, the lines extending to the left (right) are from simulations started from a random (unit) configuration. The PCAC mass jumps from a positive to a negative value across the transition, implying that the chiral limit cannot be reached in Region I.

In the right panel of Fig.~\ref{fig:1storder}, we show the same observables for $\beta=5.5$. As the transition line is continuous at this weaker coupling, we observe no jumps in the measured observables. The PCAC mass smoothly approaches zero, becomes negative before its slope changes sign around $m_0\approx-1.42$ to increase again until it vanishes for a second smaller value of the bare mass. 

In Fig.~\ref{fig:mvkmps} we show how the ratio $m_V/m_{PS}$ changes when moving from Region I (the strong coupling phase) to Region II (the weak coupling phase). In the strong coupling phase (left panel of the figure) the ratio clearly increases towards the chiral limit, which is consistent with the expectation of chiral symmetry breaking. This is not surprising, as lattice artefacts always break chiral symmetry at strong coupling. When moving towards weaker couplings into Region II (right panel) we observe that the ratio is almost constant towards the chiral limit i.e. the two states remain almost degenerate over the entire range of quark masses investigated here. This behavior, if persisting to arbitrarily small quark masses and weak couplings, would indicate the absence of spontaneous chiral symmetry breaking and it would be consistent with the expected hyperscaling behavior in an infrared conformal model \cite{DelDebbio:2010hu,DelDebbio:2010jy,DelDebbio:2010ze}. This drastic change shows a sharp transition between the bulk phase and the weak coupling phase.

\begin{figure}[t]
 \centering
 \includegraphics[scale=0.54,trim={6mm 0 6mm 0},clip]{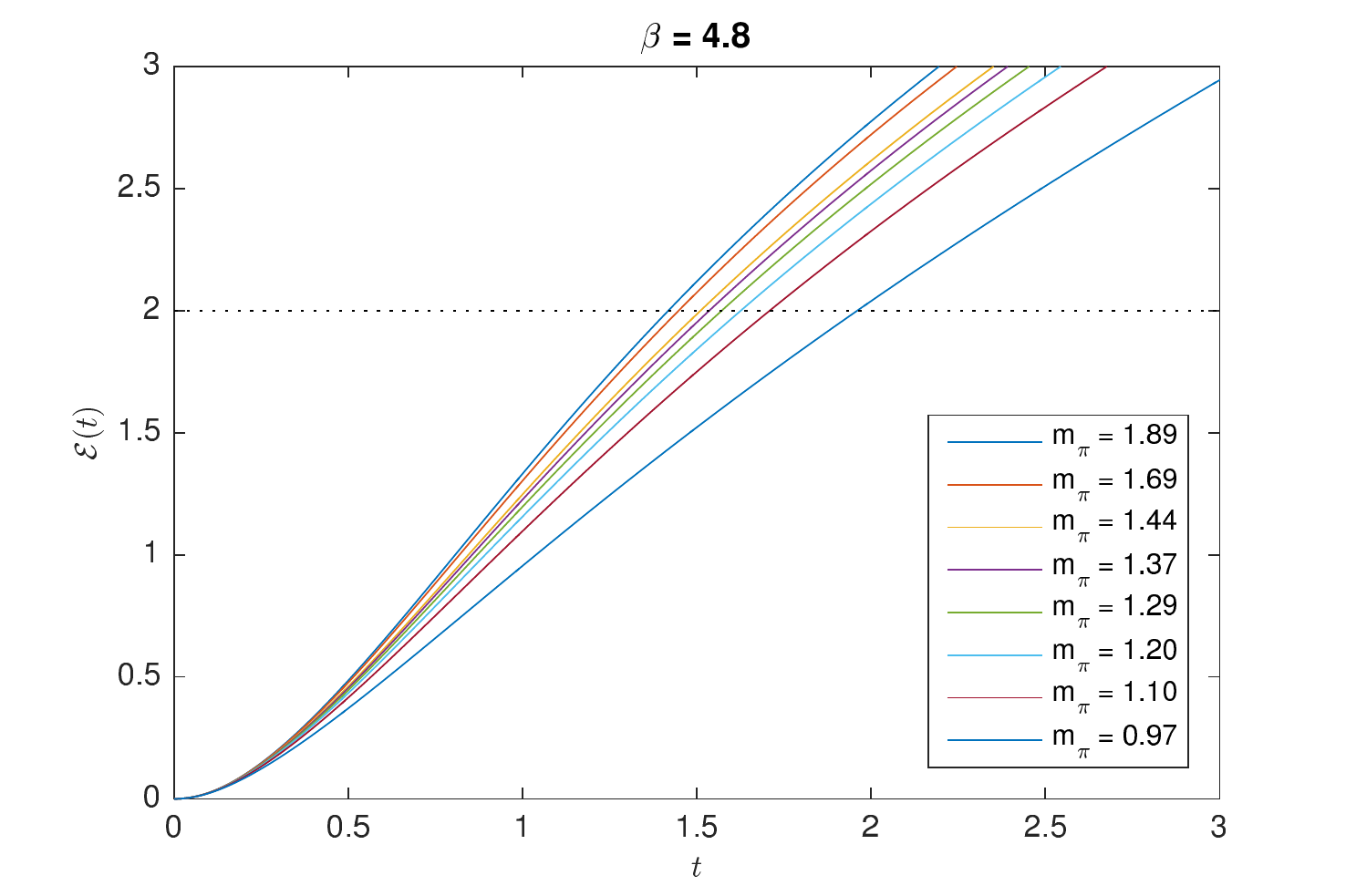}
 \includegraphics[scale=0.54,trim={6mm 0 6mm 0},clip]{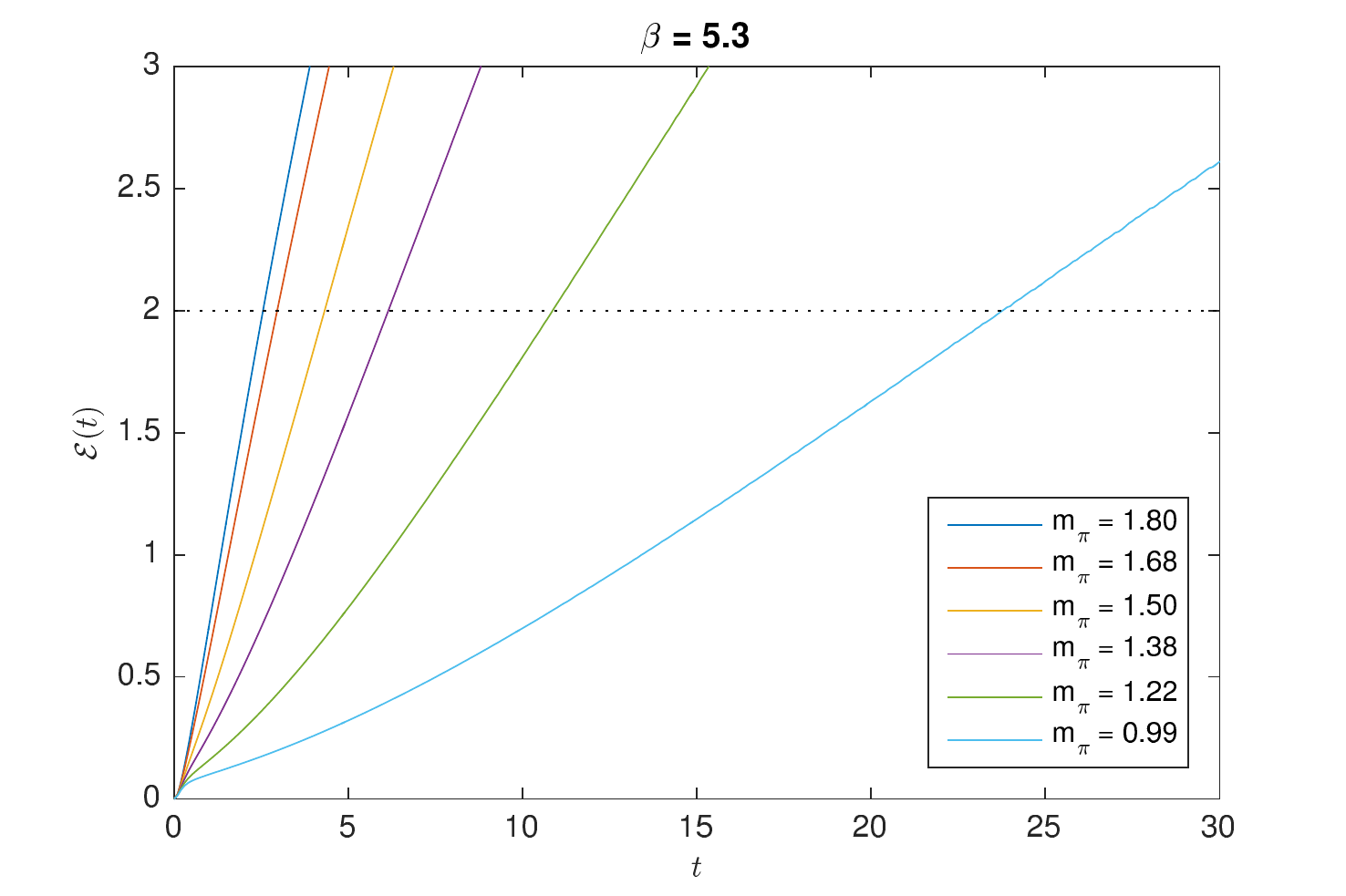}
 \caption{\footnotesize Here we show the quantity $\mathcal{E}(t)$ for two different bare couplings. The plot at $\beta=4.8$ is in the strong coupling region (Region I) and the plot at $\beta=5.3$ is in the weak coupling region (Region II). It is evident that the observable has a weak quark-mass dependence in Region I, but a strong quark-mass dependence in Region II. The dashed horizontal line is the reference value $\mathcal{E}_\mathrm{ref}=2$.}
 \label{fig:flow}
\end{figure}

Because our scan of the parameter space includes many different values of the bare coupling, in order to understand how the lattice spacing changes, we measured the scale setting observables commonly denoted by $t_0$ and $w_0$ as defined in \cite{Luscher:2010iy,Borsanyi:2012zs}. The two parameters $t_0$ and $w_0$ are defined through the quantities $\mathcal{E}(t)$ and $\mathcal{W}(t)=t\mathcal{E}'(t)$ respectively at a given reference value $\mathcal{E}(t_0) = \mathcal{E}_\mathrm{ref}$ and $\mathcal{W}(w_0^2) = \mathcal{W}_\mathrm{ref}$. The reference values used in this work are $\mathcal{E}_\mathrm{ref}=2$ and $\mathcal{W}_\mathrm{ref}=1$. In Fig.~\ref{fig:flow} we show $\mathcal{E}(t)$ for a range of bare masses at $\beta=4.8$ in the strong coupling phase (left panel) and $\beta=5.3$ in the weak coupling phase (right panel). By comparison, it is evident that, while $\mathcal{E}(t)$ shows only a weak quark-mass dependence in the strong coupling region, at weak coupling this quantity shows a very strong quark-mass dependence and, in fact, it seems to diverge in the chiral limit.

\begin{figure}[t]
 \centering
 \includegraphics[scale=0.54,trim={6mm 0 6mm 0},clip]{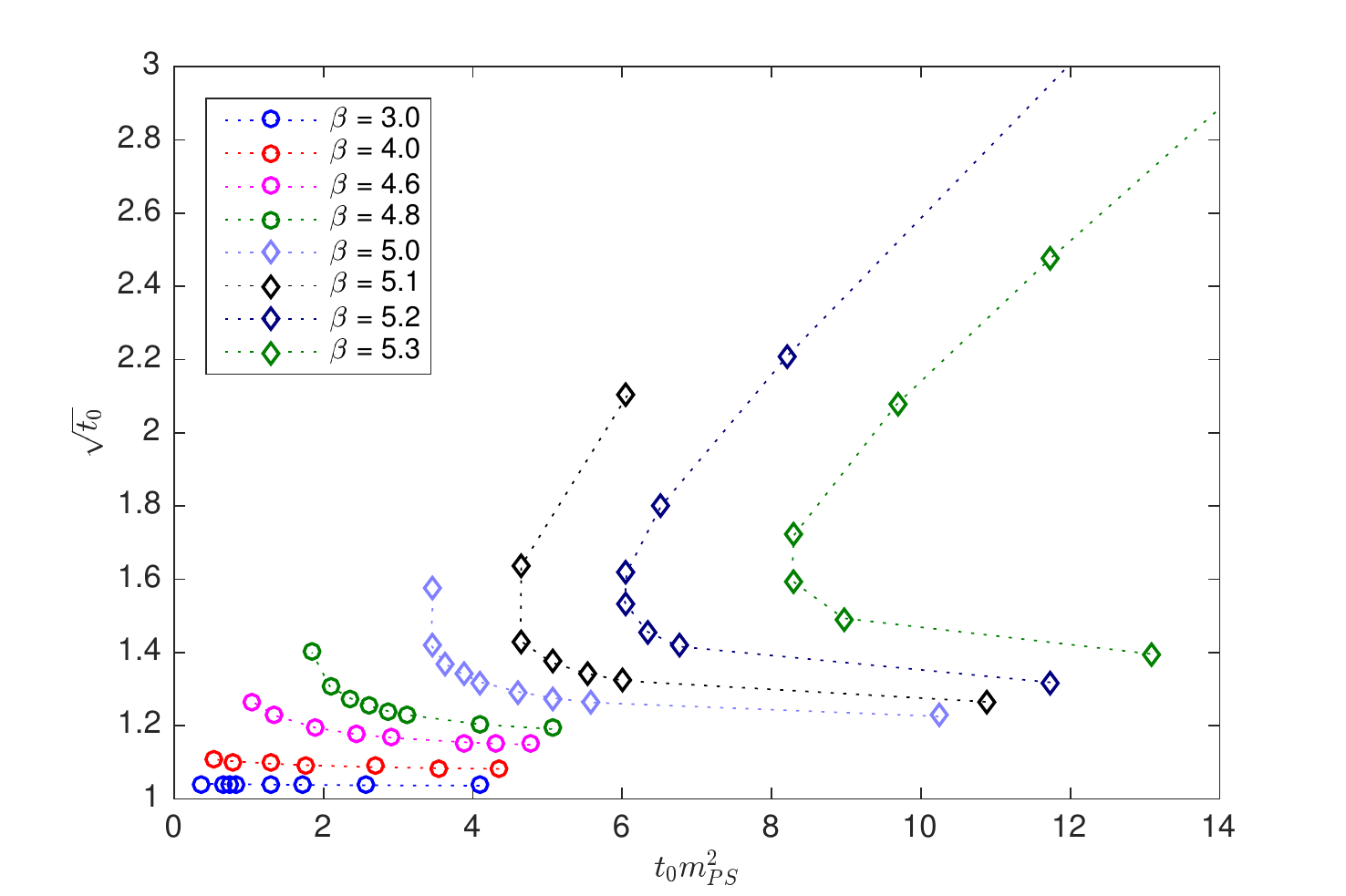}
 \includegraphics[scale=0.54,trim={6mm 0 6mm 0},clip]{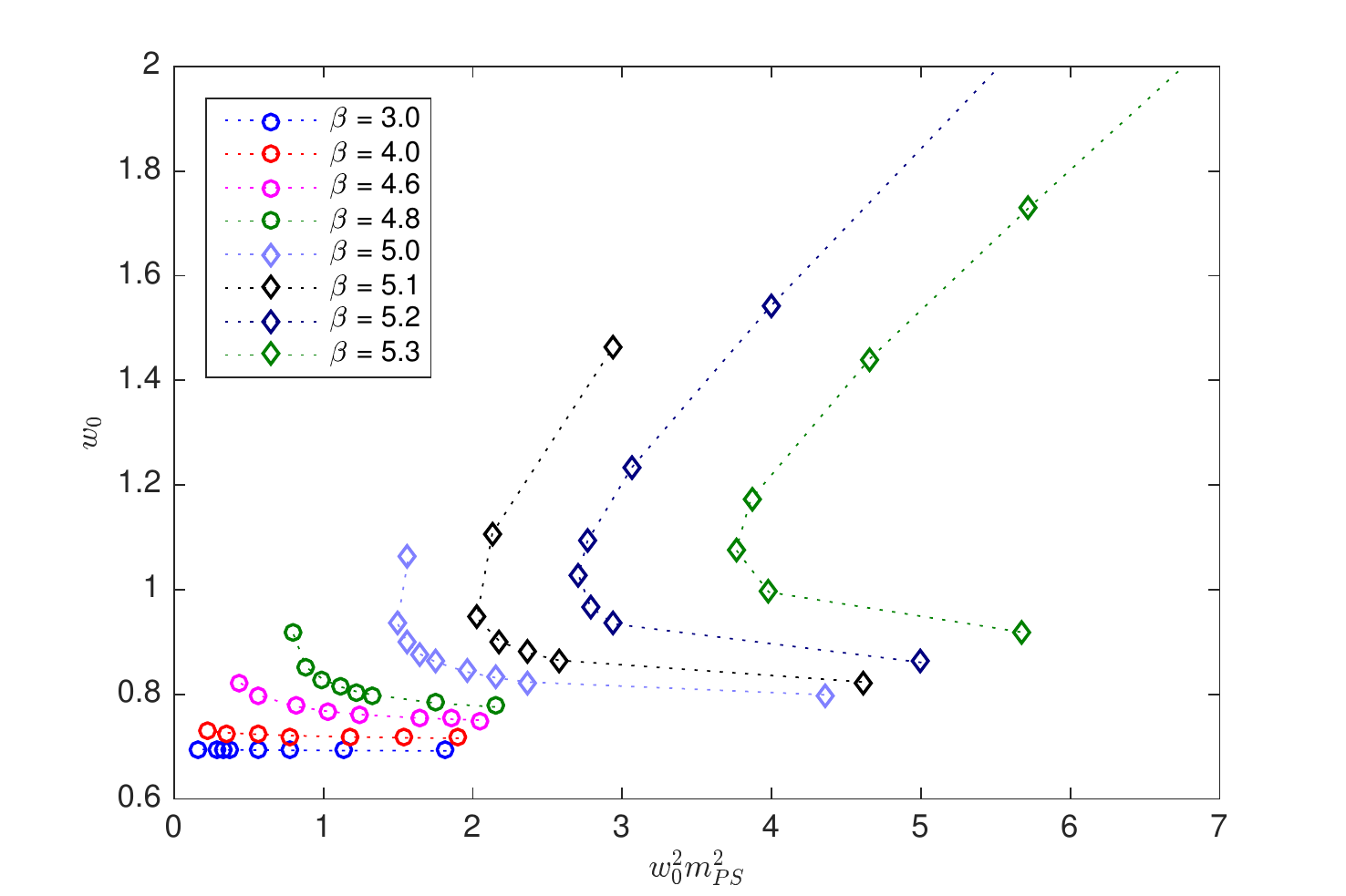}
 \caption{\footnotesize Behavior of $t_0$ and $w_0$ as a function of the lattice-spacing independent quantities $t_0m_{PS}^2$ and $w_0^2m_{PS}^2$ respectively. The observed turnaround is caused by the fact that, in the weak coupling region, both $t_0$ and $w_0$ diverges in the chiral limit.}
 \label{fig:w0t0mps}
\end{figure}

This behavior is shown in Fig.~\ref{fig:w0t0mps} where we plot $\sqrt{t_0}$ and $w_0$ as a function of $t_0m_{PS}^2$ (left panel) and $w_0^2m_{PS}^2$ (right panel)\footnote{In the published version of the proceeding, the plot in the right panel of Fig.~\ref{fig:w0t0mps} is wrong. While the label on the x-axis correctly showed $w_0^2m_{PS}^2$ the actual value plotted was $w_0m_{PS}^2$.}. In the strong coupling region the extrapolation towards the chiral limit is mild, as expected from previous studies in QCD. In contrast, in the weak coupling phase the strong quark-mass dependence of the two quantities $w_0$ and $t_0$ is clearly seen as a turnaround of the curves in Fig.~\ref{fig:w0t0mps} for $\beta\gtrsim5.0$. Assuming the model is conformal in the weak coupling region, from naive dimensional analysis and the hyperscaling relations, one would expect $t_0$ and $w_0^2$ to scale inversely with the pion mass. However, as seen on Fig.~\ref{fig:w0t0mps} the two quantities diverge much faster than what is naively expected.

\vspace{-2mm}\section{Conclusion}\vspace{-2mm}
In this work we have investigated the phase structure of the SU(3) ``sextet'' model discretized on a lattice using unimproved Wilson fermions. We extended our previous simulations performed at weak coupling by performing a complete scan of the bare parameter space. The resulting phase diagram, shown in Fig.~\ref{fig:phase_diagram}, has four interesting regions. To understand the physical behavior of the model, we have investigated the PCAC mass, pseudoscalar mass, vector mass and the scale-setting observables $w_0$ and $t_0$ as functions of the bare quark mass $m_0$ in the different regions of the phase space.

Our results show a sharp change in the qualitative behavior of the spectral and scale-setting quantities. While at strong coupling, the observations are compatible with a chirally broken model, in the weak coupling phase we fail to observe any clear indications of spontaneous chiral symmetry breaking. In particular in the weak coupling phase: the chiral transition line appears to be a continuous transition line, as opposed to a first order transition at strong coupling; the pseudoscalar and vector states remain almost degenerate over the full range of explored quark masses with the PCAC mass changing by a factor~$\sim 10$; the scale-setting observables $t_0$ and $w_0$ seem to diverge in the chiral limit at weak coupling. It is worth to stress that we have checked that the observed behavior persists at large volumes e.g. in our previous weak coupling simulations.

Despite the evidence presented, we note that we are not yet able to provide a definite answer to the question of the infrared conformality of this model. This will require the use of our data at weak coupling on the spectrum of the model, at several lattice spacings, to show consistently the presence or not of a critical behavior in the chiral limit. 

Finally the ``sextet'' model investigated here has already been extensively studied using an improved staggered fermion formulation \cite{Fodor:2012ty}. Although also in this case a conclusive answer on the infrared properties of the model has not been obtained yet, these studies indicate a preference for the model being chirally broken. Here we remark that there is a striking difference in the qualitative behavior of the spectral quantities between the two formulations e.g. the quasi degeneracy of the pseudoscalar and vector states observed here is not present in the staggered fermion formulation. 

In the future, we plan to continue our investigation of the model to clarify the infrared dynamics of the model and the observed differences with the staggered formulation.

\end{document}